\documentclass[aps,pra,reprint]{revtex4-2}

\usepackage[utf8]{inputenc}
\usepackage[T1]{fontenc}
\usepackage{amsmath,amssymb,amsfonts,bm}
\usepackage{graphicx}
\usepackage{physics}
\usepackage{mathtools}
\usepackage{float}
\usepackage{xcolor}
\usepackage{hyperref}
\usepackage{booktabs}
\usepackage{multirow}
\usepackage{subcaption}

\hypersetup{
    colorlinks=true,
    linkcolor=blue,
    citecolor=blue,
    urlcolor=blue
}

\begin{document}
\title{Interaction-Enhanced Ergotropy in Phase-Driven\\
Andreev Bound State Quantum Batteries}

\author{Disha Verma}
\email{vermadisha785@gmail.com}
\affiliation{Department of Physics, National Institute of Technology,
Tiruchirappalli 620015, India}

\author{B. Vigneshwar}
\affiliation{Department of Physics, National Institute of Technology,
Tiruchirappalli 620015, India}

\author{R. Sankaranarayanan}
\affiliation{Department of Physics, National Institute of Technology,
Tiruchirappalli 620015, India}
\date{\today}

\begin{abstract}

We investigate a phase-driven quantum battery composed of two interacting Andreev bound state (ABS) units, providing a minimal superconducting platform for coherent energy storage. By analyzing the ergotropy dynamics under a superconducting phase ramp, we show that the interplay between avoided-crossing excitation and interaction-induced hybridization strongly modifies the charging process. In the high-transparency regime relevant for graphene SNS junctions, the interaction enhances the stored extractable work and generates pronounced oscillatory charging dynamics associated with coherent redistribution between coupled ABS sectors. The phase-resolved evolution further reveals optimal charging windows during the Josephson cycle, indicating the possibility of phase-programmable energy extraction through partial-cycle operation. Overall, our results identify interaction-assisted avoided-crossing dynamics as a microscopic mechanism for controllable energy storage in superconducting quantum batteries.

\end{abstract}

\maketitle

\section{Introduction}

Quantum batteries (QBs) are quantum systems designed to store, manipulate, and extract energy using coherent quantum dynamics~\cite{Allahverdyan2004,Binder2015,Campaioli2018}.  Theoretical studies have proposed a variety of QB models, including spin systems \cite{le2018spin}, cavity-QED systems \cite{dou2022cavity}, Gaussian and free-fermion batteries \cite{gemme2023off, catalano2024frustrating}, dissipative charging protocols \cite{cavaliere2025gaussian, grazi2025fermion, massa2025collisional}, topological quantum systems \cite{lu2025topological} and many-body interacting setups \cite{ cavaliere2025blockade, grazi2026quenches, chand2026spin, farina2026phase, pavone2026cluster, grazi2025finite}. A central question in this field is how quantum coherence, interactions, and spectral structure influence both the amount of extractable work (ergotropy) and the rate at which energy can be deposited. Early studies have focused on spin-based models~\cite{grazi2024controlling,catalano2024frustrating,verma2025dynamics} and collective charging protocols, including central-spin, Dicke, and XY-chain models where many spins collectively couple to a field or to each other, showing strong speedups and phase-transition–controlled storage ~\cite{Alicki2013,Campaioli2017,Yang2023Three-level,Downing2023A,Ahmadi2024Nonreciprocal}. Superconducting qutrit and coupled transmon qubits have been used or proposed experimentally, demonstrating optimized charging, self-discharge behavior, and coherence-assisted efficiency \cite{Crescente2020Charging}.  More recently, attention has extended to physically motivated platforms, including superconducting circuits, cavity-QED systems, and hybrid graphene-based structures, where both coherent and dissipative mechanisms of energy storage have been explored~\cite{Blais2004,Krantz2019,verma2026ergotropy}.
Among these, superconducting hybrid structures are particularly promising due to their high degree of controllability via phase bias, gate voltages, and junction transparency \cite{elghaayda2025performance, dou2023superconducting}. These systems naturally provide a tunable energy spectrum, enabling alternative charging paradigms beyond conventional time-dependent driving.

While a wide variety of quantum battery platforms have been proposed, most existing models rely on externally driven Hamiltonians to induce charging dynamics. An alternative approach is to exploit systems whose energy spectrum is intrinsically controllable, allowing energy storage to emerge directly from tunable spectral properties rather than engineered driving pulses. In this context, superconductor--normal--superconductor (SNS) junctions hosting Andreev bound states (ABS) provide a particularly attractive platform. Unlike conventional quantum batteries based on explicit time-dependent driving fields, ABS systems possess a phase-controlled spectrum in which the energy levels depend directly on the superconducting phase difference across the junction~\cite{Beenakker1991,Beenakker2006}. As a result, coherent energy injection can be achieved through phase modulation alone.

A central feature of the ABS spectrum is the avoided crossing near $\phi=\pi$, where the excitation gap becomes minimal. This region naturally enables non- adiabatic excitation during a phase sweep, allowing efficient charging without requiring complicated pulse engineering or many-body control protocols. In this sense, ABS systems establish a distinct paradigm of quantum batteries based on Hamiltonian engineering through phase-controlled superconducting spectra.

Graphene-based SNS junctions are especially promising in this regard due to their high transparency, gate-tunable carrier density, and multiple transport channels associated with spin and valley degrees of freedom~\cite{Bretheau2017,verma2026ergotropy}. Owing to Klein tunneling and ballistic transport \cite{Du2008}, graphene Josephson junctions can operate close to the nearly transparent regime, where the ABS spectrum becomes strongly phase sensitive and the avoided-crossing physics is particularly pronounced. These features provide a versatile platform for controllable superconducting energy storage.

Based on these findings, the central question addressed in this work is whether superconducting phase dynamics and interaction-induced hybridization can be exploited as controllable resources for quantum energy storage. More specifically, we ask whether a coupled ABS system can generate useful extractable work through purely phase-driven dynamics, and how the inter-junction interaction modifies the charging process.

Unlike conventional quantum batteries based on externally driven fields, the present platform enables charging through direct manipulation of the superconducting phase-dependent spectrum. This naturally raises several important questions: Can avoided-crossing dynamics generate efficient nonadiabatic charging? Does coherent coupling between ABS sectors enhance the stored ergotropy? And can the charging process be optimized through partial Josephson phase cycles rather than requiring a full phase sweep?

To address these questions, we investigate a minimal interacting quantum battery composed of two coupled ABS units. Motivated by recent experiments demonstrating coherent coupling between ABS qubits in superconducting hybrid devices~\cite{PitaVidal2024,Zazunov2003}, we analyze the charging dynamics induced by a finite superconducting phase ramp and quantify the extractable work using ergotropy.
In particular, we focus on the interplay between interaction-induced hybridization and phase-controlled dynamics in the high-transparency regime relevant for graphene SNS junctions. Our goal is to understand how coherent coupling modifies the generation, redistribution, and extraction of useful work in interacting ABS quantum batteries.

%%%%%%%%%%%%%%%%%%%%%%%%%%%%%%%%%%%%%%%%%%%%%%%%%%%%%%%%%%%%%%%%%%%%%%%%%%%%%%%%%%%%%%%%%%%%%%%%%%%%%%%%%%%%%%%%%%%%%%%%%%%%%%%%%%%%%%%%%%%%%%%%%%%%
\section{Model}

We model each Andreev bound state (ABS) unit as an effective two-level system arising from the subgap spectrum of a short superconductor--normal--superconductor (SNS) junction \cite{park2024controllable}, and construct a minimal quantum battery from two such units coupled via an inter-junction interaction. The total Hamiltonian is written as
\begin{equation}
H(t)=H_{\mathrm{ABS}}(t)+H_{\mathrm{int}}(J),
\label{total}
\end{equation}
The two ABS units constitute the quantum battery degrees of freedom, while the externally controlled superconducting phase difference acts as the charging resource. The interaction term does not directly contribute to the stored energy but facilitates coherent population transfer and correlation generation between the two ABS units. Consequently, the extractable work is evaluated with respect to a battery Hamiltonian associated with the instantaneous ABS energy spectrum. The phase-driven dynamics of the coupled ABS system are governed by the Hamiltonian

\begin{equation}
\begin{split}
H_{\mathrm{ABS}}(t)
={}&
\Delta\cos\!\left[\frac{\phi(t)}{2}\right]
\left(\sigma_z^{(1)}+\sigma_z^{(2)}\right) \\
&+
\Delta\sqrt{1-T}\,
\sin\!\left[\frac{\phi(t)}{2}\right]
\left(\sigma_x^{(1)}+\sigma_x^{(2)}\right).
\end{split}
\label{eq2}
\end{equation}
written in a fixed pseudospin (computational) basis. Here, $\Delta$ denotes the superconducting gap, $T$ is the normal-state transmission probability (junction transparency) of the weak link $(0\le T\le1)$, and $\sigma_\alpha^{(i)}$ are the Pauli operators acting on the (i)th ABS unit. Diagonalization of Eq.~(\ref{eq2}) yields the phase-dependent ABS energy levels, which define the instantaneous energy-storage sector used in the evaluation of ergotropy.

The superconducting phase is driven according to
\begin{equation}
\phi(t)=
\begin{cases}
\phi_0+vt, & t \le \tau_{\mathrm{ramp}},\\
\phi_f, & t > \tau_{\mathrm{ramp}},
\end{cases}
\label{FI}
\end{equation}
where $\tau_{\mathrm{ramp}}=(\phi_f-\phi_0)/v$.  where \(v=d\phi/dt\) is the phase-ramp velocity. Through the Josephson relation
\(
d\phi/dt = 2eV/\hbar,
\) the ramp velocity is directly controlled by the applied voltage bias \(V\) across the junction. The charging performance is evaluated at the end of the ramp, $t=\tau_{\mathrm{ramp}}$ (corresponding to $\phi=\phi_f=2\pi$).  In this sense, the phase ramp injects energy into the ABS sector by driving the system through its phase-dependent spectrum. The corresponding Andreev bound-state (ABS) energy levels are given by $\pm E_A(\phi)$:

\begin{equation}
E_A(\phi)=\Delta\sqrt{1-T\sin^2\left(\frac{\phi}{2}\right)},
\label{ea}
\end{equation}
which exhibits a minimum at $(\phi=\pi)$. For finite transparency ($T<1$), this corresponds to an avoided crossing with a minimal gap  $\Delta_{\mathrm{gap}}=2\Delta\sqrt{1-T}$, which plays a central role in the charging dynamics.
The phase $\phi(t)$ acts as an external control parameter and therefore serves as the charging protocol, while the two coupled ABS units constitute the battery degrees of freedom. 

This effective description can be understood as a low-energy projection of the Bogoliubov--de Gennes (BdG) Hamiltonian of an SNS junction, where the subgap spectrum consists of a particle--hole symmetric pair of ABS at energies $\pm E_A(\phi)$. In the short-junction limit ($L\ll\xi$), and for dynamics restricted to the subgap sector, a two-level truncation is justified since continuum states lie at $|E|\ge\Delta$ \cite{Bretheau2017}. In the instantaneous ABS eigenbasis, each unit is described by $H_{\mathrm{ABS}}(\phi)=E_A(\phi)\sigma_z$, while in a fixed pseudospin basis the Hamiltonian containing both longitudinal ($\sigma_z$) and transverse ($\sigma_x$) components. The transverse term phenomenologically captures the mixing between Andreev levels induced by finite transmission and nonadiabatic phase dynamics. The minimal gap governs the nonadiabatic dynamics during the phase sweep. 

The interaction term
\begin{equation}
H_{\mathrm{int}}(J)
=
J\,\sigma_x^{(1)}\sigma_x^{(2)}
\label{int}
\end{equation}
is introduced as a minimal effective description of coherent coupling between the two ABS units, where $J$ denotes the interaction strength. The transverse form of the coupling phenomenologically captures interaction-induced hybridization and coherent population redistribution within the low-energy ABS subspace. Such effective coupling may arise from overlapping ABS wave functions, shared superconducting elements, or circuit-mediated interactions between nearby junctions~\cite{Blais2004,Janvier2015,PitaVidal2024}. In the present work, we treat $J$ as an effective phenomenological parameter and focus on its influence on the charging dynamics and ergotropy generation.
The operators $\sigma_x$ and $\sigma_z$ are defined with respect to a fixed computational 
basis, while the phase dependence is incorporated through time-dependent coefficients. 
This avoids a time-dependent basis transformation and ensures a consistent definition 
of the energy eigenbasis and ergotropy throughout the charging protocol.

%%%%%%%%%%%%%%%%%%%%%%%%%%%%%%%%%%%%%%%%%%%%%%%%%%%%%%%%%%%%%%%%%%%%%%%%%%%%%%%%%%%%%%%%%%%%%%%%%
\section{Charging protocol and ergotropy}
%%%%%%%%%%%%%%%%%%%%%%%%%%%%%%%%%%%%%%%%%%%%%%%%%%%%%%%%%%%%%%%%%%%

While the Hamiltonian in Eq. \ref{eq2}  is expressed in a fixed pseudospin (computational) basis and governs the phase-driven dynamics of the ABS system, the evaluation of extractable work is performed with respect to the instantaneous ABS energy spectrum. We therefore introduce the battery Hamiltonian in the instantaneous ABS eigenbasis, where the phase-dependent ABS levels are characterized by the eigenenergies $(\pm E_A(\phi))$. The battery Hamiltonian is taken to be

\begin{equation}
H_B(\phi)=E_A(\phi)\bigl(\sigma_z^{(1)}+\sigma_z^{(2)}\bigr),
\label{battery}
\end{equation}
which defines the instantaneous ABS energy sector and serves as the reference Hamiltonian for the calculation of ergotropy.

 Taking $\phi(0)=0$ the system is initialized in the ground state of the battery Hamiltonian $H_B(0)$. To quantify the extractable work stored in the system, we employ the concept of ergotropy~\cite{Allahverdyan2004,Binder2015,Campaioli2017}, defined as
\begin{equation}
\mathcal{W}(t)=\Tr[\rho(t) H_B]-\Tr[\rho_{\mathrm{p}}H_B],
\label{eq:ergotropy}
\end{equation}
where $\rho(t)$ is the time evolved system density matrix and $H_B$ denotes the battery Hamiltonian. The passive state $\rho_{\mathrm{p}}$ is constructed by rearranging the eigenvalues of $\rho(t)$ in decreasing order onto the eigenvectors of $H_B$ ordered by increasing energy.

The instantaneous ergotropy is evaluated using eq.(\ref{eq:ergotropy})  which measures the useful energy stored relative to the instantaneous ABS energy spectrum. Since the system evolves under a Hamiltonian $H(t)$ that differs from $H_B$, the state $\rho(t)$ is generally not diagonal in the energy eigenbasis, giving rise to non-passivity and finite ergotropy.

In the present work, we primarily focus on the final ergotropy,
\begin{equation}
W_{\mathrm{final}}
=
\mathcal{W}(\tau_{\mathrm{ramp}}),
\label{work}
\end{equation}
which quantifies the extractable work stored in the ABS sector after completion of the full
\[
0\rightarrow2\pi
\]
phase sweep. In addition to the final stored work, we also analyze the phase-resolved ergotropy dynamics during the charging evolution in order to identify the microscopic mechanisms responsible for interaction-enhanced energy storage.
Since the ABS spectrum itself depends on the superconducting phase, the battery Hamiltonian is intrinsically time dependent. The instantaneous ergotropy therefore quantifies the extractable work available relative to the instantaneous ABS energy basis and directly measures the degree of non-passivity generated during the phase-driven evolution.

Importantly, the ergotropy is distinct from the total external work injected by the phase source. Although the superconducting phase ramp continuously transfers energy into the system, only a fraction of this energy becomes operationally extractable as useful work. The ergotropy therefore characterizes the recoverable part of the injected energy and provides an upper bound on the useful work that can be extracted from the charged ABS sector through unitary operations.

%%%%%%%%%%%%%%%%%%%%%%%%%%%%%%%%%%%%%%%%%%%%%%%%%%%%%%%%%%%%%%%%%

\section{Results}
The numerical simulations are performed using a phase-driven protocol in which the superconducting phase is linearly ramped from $\phi_0=0$ to $\phi_f=2\pi$. This choice corresponds to a complete Josephson cycle, ensuring that the system traverses the avoided crossing in the ABS spectrum near $\phi \approx \pi$, which governs the nonadiabatic energy transfer during charging~\cite{Zazunov2003,Shevchenko2010}.
We work in units where $\hbar=1$ and energies are normalized by $\Delta$, so that time is measured in units of $\hbar/\Delta$. The quantity $v/\Delta_{\mathrm{gap}}^2$ serves as a convenient scaling variable that compares the phase ramp rate to the square of the minimum gap, which plays a central role in nonadiabatic transitions near the avoided crossing. This normalization highlights how the charging dynamics depend on the relative strength of driving and the intrinsic spectral scale.

Unless otherwise stated, we set $\Delta=1$ to fix the energy scale. We focus on the high-transparency regime, taking $T=0.98$ for representative results, which is experimentally relevant for graphene and superconducting weak links and produces a small avoided-crossing gap in the ABS spectrum, thereby enhancing nonadiabatic transitions\cite{PitaVidal2024}. The ramp speed is chosen as $v = 5.0$. This choice avoids both the nearly adiabatic regime, where excitation generation is weak, and the strongly diabatic regime, where coherent redistribution between ABS states becomes inefficient.
In this regime, Landau–Zener transitions across the avoided crossing at $\phi \approx \pi$ 
are significant but not fully diabatic, enabling efficient population transfer while 
preserving coherent dynamics~\cite{Shevchenko2010}. The role of the avoided crossing near $\phi=\pi$ can be understood from the Landau--Zener picture. 
For an isolated avoided crossing under a linear phase sweep, the nonadiabatic transition probability is approximately controlled by \cite{verma6659398phase}
\begin{equation}
P_{\rm LZ}
\sim
\exp\left[
-\alpha
\frac{\Delta_{\rm gap}^2}{v}
\right],
\end{equation}
where $\alpha$ is a model-dependent numerical factor.  The scaling variable $v/\Delta_{\mathrm{gap}}^2$ therefore provides a natural dimensionless measure of adiabaticity in the charging protocol. Thus, increasing the transparency reduces the avoided-crossing gap and enhances nonadiabatic population transfer. However, the final ergotropy is not determined by $P_{\rm LZ}$ alone; it also depends on how the generated excitations are redistributed among the coupled ABS states. The interaction $J$ therefore provides an additional control knob by modifying the effective hybridization pathways after the avoided crossing.

The chosen interaction range spans from weakly coupled ABS units to the strongly hybridized regime, where interaction-induced hybridization significantly modifies the ABS-level structure and charging dynamics.

%%%%%%%%%%%%%%%%%%%%%%%%%%%%%%%%%%%%%%%%%%%%%%%%%%%%%%%%%%
%%%%%%%%
\subsection{Interaction-assisted charging regimes in the high-transparency limit}

To investigate the role of inter-junction coupling in the phase-driven charging process, we analyze the final ergotropy as a function of the scaled interaction strength
\begin{equation}
\frac{J}{E_A(\phi_f)},
\end{equation}
where $E_A(\phi_f)$ is the characteristic ABS energy scale evaluated at the final phase $\phi_f=2\pi$. Since $\sin^2(\phi_f/2)=0$, one has (eq. \ref{ea})
\begin{equation}
E_A(\phi_f)=\Delta,
\end{equation}
so that the scaled interaction reduces to $J/\Delta$.
The final ergotropy
\begin{equation}
W_{\mathrm{final}} = \mathcal{W}(\phi_f=2\pi)
\end{equation}
quantifies the amount of extractable work stored in the battery after completion of the phase ramp. Throughout this section, the ramp speed is fixed to $v=5.0$.

\begin{figure}[t]
\centering
\includegraphics[width=0.72\columnwidth]{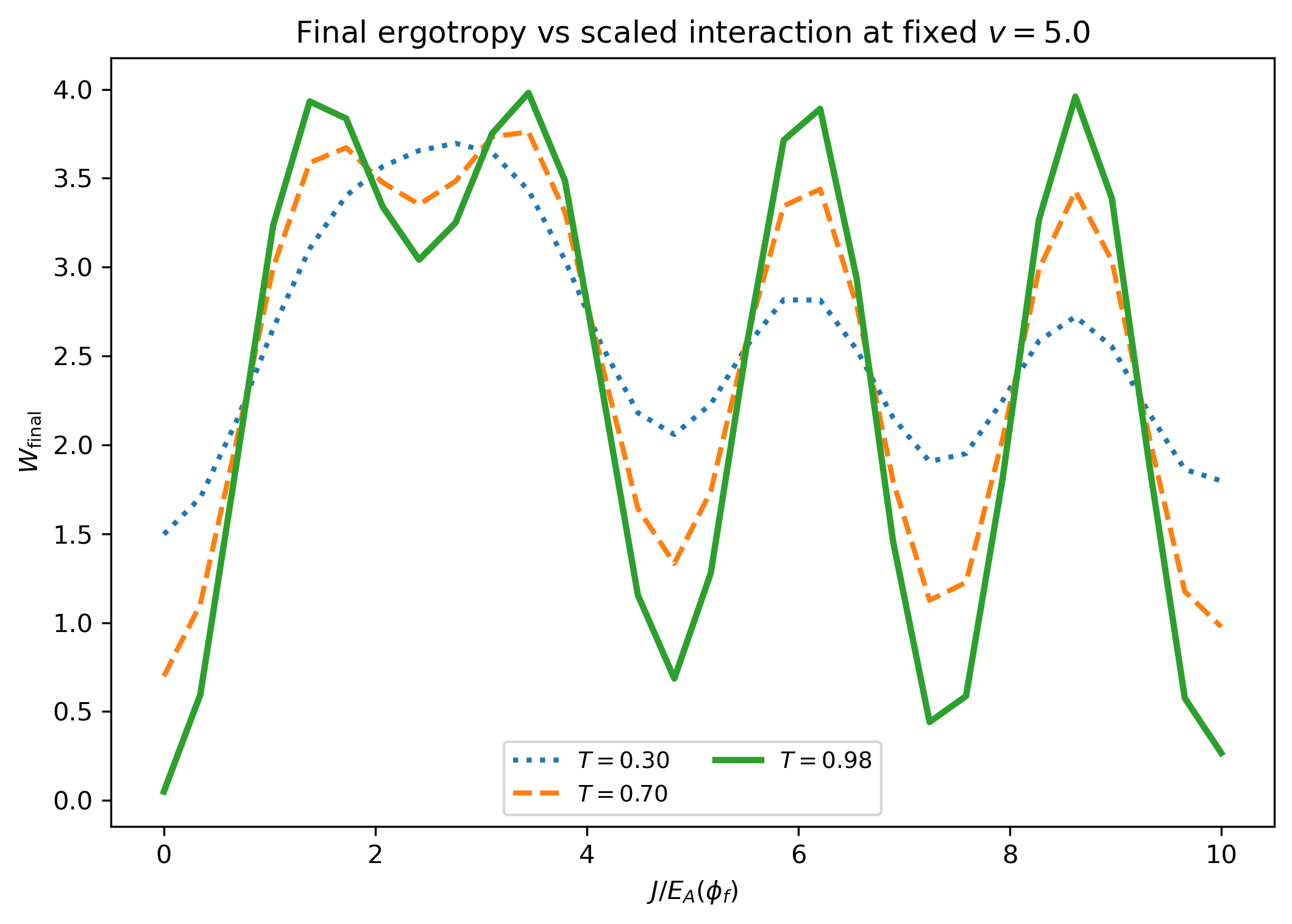}
\caption{
Final ergotropy $W_{\mathrm{final}}$ as a function of the scaled interaction strength $J/E_A(\phi_f)$ at fixed ramp speed $v=5.0$ for representative transparencies $T=0.30$ (dotted), $0.70$ (dashed), and $0.98$ (solid). The interaction initially enhances the stored extractable work, while larger interaction strengths produce a pronounced oscillatory regime associated with coherent interaction-driven redistribution dynamics. The strongest modulation occurs in the nearly ballistic regime $T=0.98$.
}
\label{fig:Wfinal_vs_Jscaled}
\end{figure}
Figure~\ref{fig:Wfinal_vs_Jscaled} reveals that the interaction strongly modifies the charging behavior of the coupled ABS quantum battery. For all transparencies, the final ergotropy initially increases rapidly with interaction strength, demonstrating that inter-junction coupling constructively assists the generation of extractable work.
This enhancement originates from interaction-induced hybridization between the two ABS sectors. The coupling enlarges the accessible set of non-passive many-body configurations during the phase evolution and thereby promotes more efficient redistribution of excitation energy generated during the ramp.

The figure further reveals three qualitatively distinct charging regimes.
In the weak-coupling regime, $J/E_A(\phi_f)\ll1$ the final ergotropy increases rapidly as the interaction is turned on. This initial enhancement occurs for all transparencies and demonstrates that even moderate inter-junction coupling can substantially improve the charging
efficiency. Physically, the interaction hybridizes the two ABS sectors and enables coherent redistribution of excitation
energy between the coupled junctions, thereby enlarging the accessible set of non-passive many-body states during
the phase ramp. The transparency dependence in this regime is also significant. For smaller transparency, the ABS spectrum remains smoother and the avoided-crossing gap $(\Delta_{gap})$ is relatively large, producing more adiabatic phase evolution.
Consequently, the charging dynamics remain comparatively stable and the interaction dependence appears smoother.
In contrast, as the transparency approaches the ballistic limit $(T\to1)$, the avoided-crossing gap becomes strongly reduced and
the dynamics near $\phi=\pi$ become increasingly nonadiabatic. This enhanced phase sensitivity allows the interaction assisted redistribution mechanism to operate more efficiently.

At intermediate interaction strength  $J/E_A(\phi_f)\approx1$, the stored extractable work reaches its largest values over a broad parameter range. In this regime, the interaction-induced hybridization and the phase-driven nonadiabatic excitation near the avoided crossing cooperate most efficiently, leading to strong population transfer into highly non-passive states.

For stronger interaction strengths  $J/E_A(\phi_f)>1$, the charging dynamics enter a coherent oscillatory regime characterized by repeated enhancement and suppression of the final ergotropy. In this regime, the interaction energy becomes comparable to or larger than the intrinsic ABS scale, generating additional coherent redistribution timescales during the phase evolution. The resulting competition between phase-driven excitation and interaction-induced redistribution produces alternating constructive and destructive interference in the stored extractable work.

The transparency dependence is particularly important. As the transparency approaches the ballistic limit, the avoided-crossing gap $\Delta_{\mathrm{gap}}$
becomes strongly reduced, enhancing the nonadiabatic response near $\phi=\pi$.
Consequently, the oscillatory interaction-driven regime becomes increasingly pronounced for larger transparency.
The strongest modulation occurs for $T=0.98$,
which is especially relevant for graphene SNS junctions operating close to the ballistic transport regime due to Klein tunneling and highly transparent Josephson channels \cite{titov2006josephson,Du2008}. . Since this regime exhibits the strongest phase sensitivity and the largest interaction-induced modulation, the following sections focus primarily on the high-transparency case
in order to investigate the microscopic phase-resolved charging dynamics in the experimentally relevant graphene limit.
The transparency comparison in Fig.~\ref{fig:Wfinal_vs_Jscaled} therefore serves two purposes. First, it shows that interaction-assisted charging is present across different transparency regimes. Second, it identifies the high-transparency limit as the most sensitive and physically relevant regime for graphene SNS junctions.

This motivates a more detailed phase-resolved analysis at fixed $T=0.98$. Since $W_{\mathrm{final}}$ only measures the stored extractable work at the end of the $0\to2\pi$ cycle, it does not reveal when during the ramp the ergotropy is generated or whether the large oscillations arise from early-cycle charging, post-crossing growth, or coherent backflow. Therefore, in the next subsection we fix $T=0.98$ and examine $W(\phi)$ during the full Josephson cycle for representative interaction strengths.

%%%%%%%%%%%%%%%%%%%%%%%%%%%%%%%%%%%%%%%%%%%%%%%%%%%%%%%%%%%%%%%%%%%%%%%%%%%%%%%%%%%%%%%
\subsection{Phase-resolved charging dynamics and optimal partial-cycle extraction}

\begin{figure}[t]
\centering

\begin{subfigure}{0.8\linewidth}
    \centering
    \includegraphics[width=\linewidth]{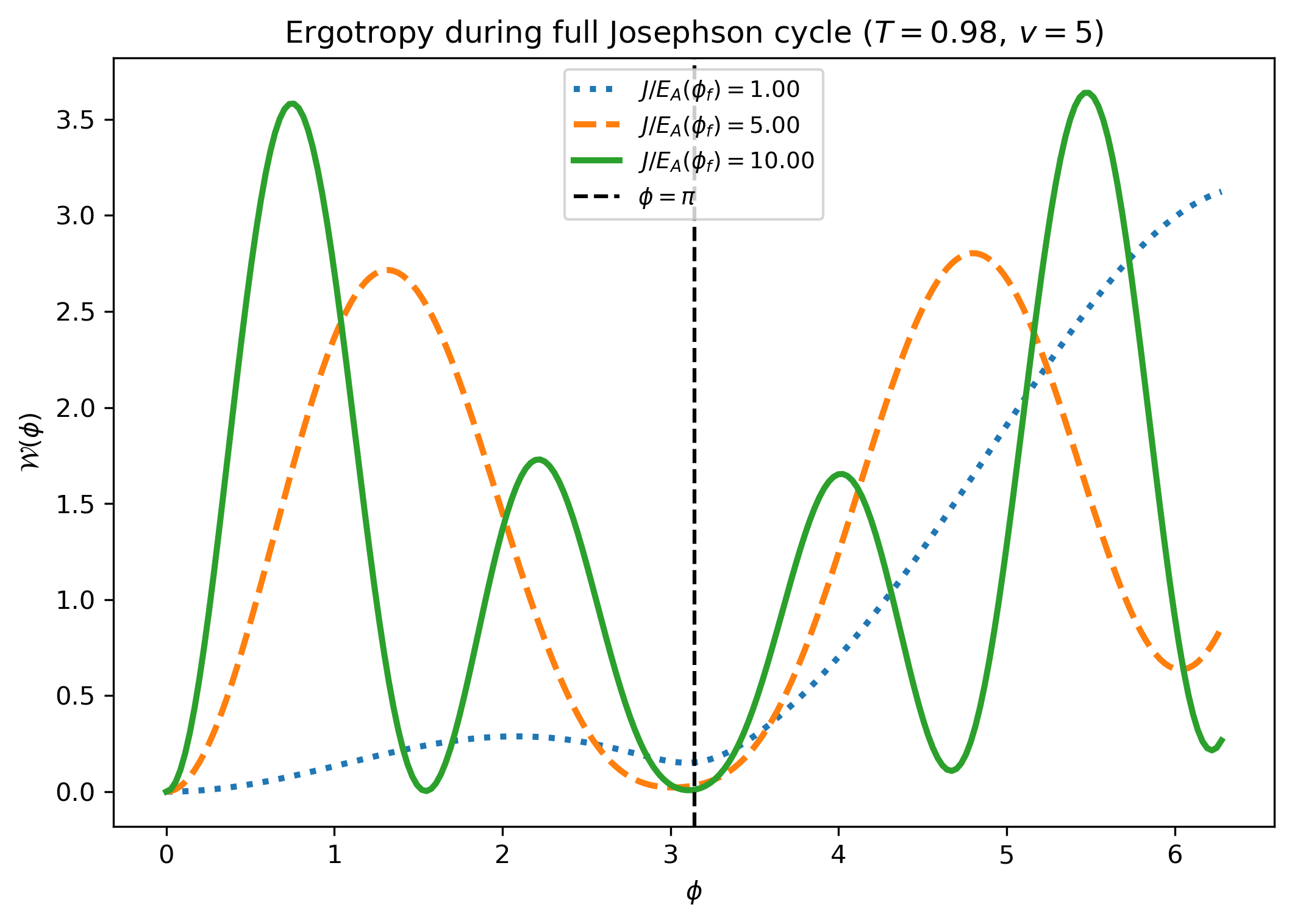}
    \caption{}
\end{subfigure}

\vspace{0.4cm}

\begin{subfigure}{0.8\linewidth}
    \centering
    \includegraphics[width=\linewidth]{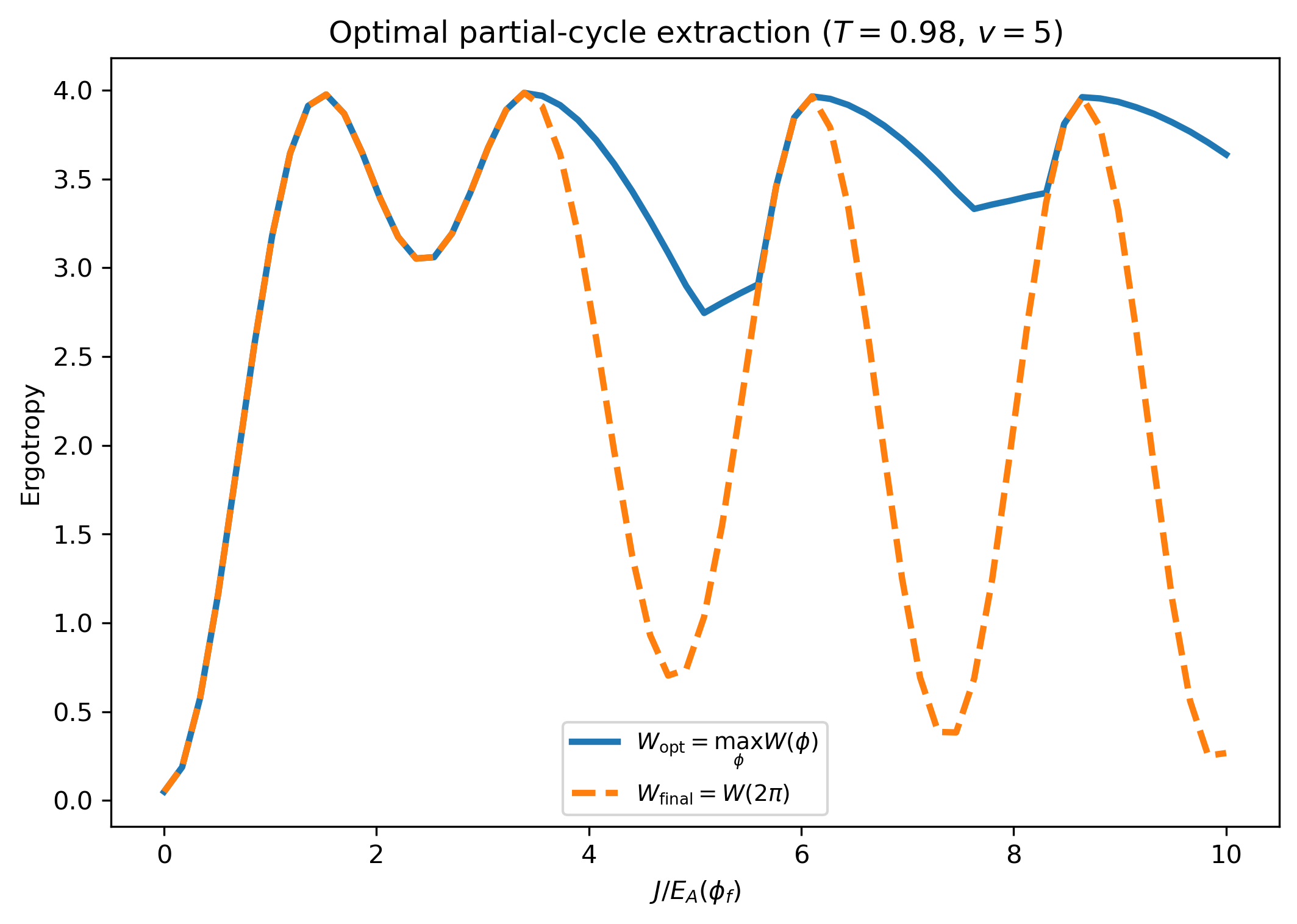}
    \caption{}
\end{subfigure}

\caption{
(a) Phase-resolved ergotropy $\mathcal{W}(\phi)$ during a full Josephson cycle for different interaction strengths at fixed transparency $T=0.98$ and ramp velocity $v=5$. The dashed vertical line marks the avoided-crossing point $\phi=\pi$. 
(b) Comparison between the optimal extractable ergotropy
\(
W_{\mathrm{opt}}=\max_{\phi}\mathcal{W}(\phi)
\)
and the final-cycle ergotropy
\(
W_{\mathrm{final}}=\mathcal{W}(2\pi)
\)
as a function of interaction strength.
}
\label{fig:partial_cycle}
\end{figure}
To resolve the microscopic origin of the interaction-assisted charging behavior, we now analyze the phase-resolved evolution of the ergotropy during the full Josephson cycle. Unlike the final-cycle analysis discussed previously, the phase-resolved dynamics directly reveal how the avoided crossing and the interaction-induced hybridization cooperate during the charging process.

Figure~\ref{fig:partial_cycle} (a) shows the evolution of the ergotropy $\mathcal{W}(\phi)$
for representative interaction strengths in the high-transparency regime.
For weak interaction strength, the charging dynamics remain relatively smooth and monotonic. The ergotropy increases predominantly after the avoided crossing near $\phi\simeq\pi$,
indicating that the main charging process is initiated by nonadiabatic excitation across the minimum ABS gap.
As the interaction strength increases, the charging dynamics become strongly oscillatory and multiple local maxima emerge during a single Josephson cycle. This behavior demonstrates that the charging process is no longer governed solely by the avoided-crossing excitation, but also by coherent redistribution dynamics generated by the interaction between the two ABS sectors.

Physically, the avoided crossing injects excitation into the coupled ABS system, while the interaction hybridization redistributes this excitation between different many-body configurations. The resulting competition between phase-driven excitation and interaction-induced redistribution generates additional coherent oscillation timescales during the phase evolution.

The local maxima of $\mathcal{W}(\phi)$
correspond to coherent alignment between these two processes, where the redistribution dynamics temporarily maximize the population imbalance associated with non-passive states. In contrast, the subsequent minima originate from coherent backflow processes that partially reduce the extractable work before the completion of the full cycle.
This behavior reveals that the charging dynamics of the interacting ABS battery are intrinsically phase selective. In particular, sufficiently strong interactions allow large ergotropy to be generated at specific phase values smaller than $2\pi$.
The charging process therefore becomes operationally programmable through the superconducting phase evolution.

The strongly oscillatory phase-resolved dynamics shown in Fig.~\ref{fig:partial_cycle}(a) suggest that completion of the full Josephson cycle is not necessarily the optimal operational strategy for extracting useful work from the interacting ABS quantum battery.
To quantify this effect, we define the optimal partial-cycle ergotropy
\[
W_{\mathrm{opt}}
=
\max_{\phi} \mathcal{W}(\phi),
\]
which measures the maximum extractable work reached at any intermediate phase during the ramp. This quantity is compared with the conventional full-cycle ergotropy
\[
W_{\mathrm{final}}
=
\mathcal{W}(2\pi),
\]
which assumes extraction only after completion of the full Josephson cycle.
Figure~\ref{fig:partial_cycle}(b) reveals that the difference between $W_{\mathrm{opt}}$ and $W_{\mathrm{final}}$ becomes increasingly pronounced at strong interaction strength. While both quantities coincide in the weak-coupling regime, large deviations emerge once coherent oscillatory dynamics develop. In this regime, the full-cycle protocol can significantly underestimate the actual charging capability of the system because coherent redistribution occurring after the optimal charging point partially reduces the stored ergotropy before the end of the cycle.

This demonstrates that optimal operation of interacting ABS quantum batteries does not necessarily require completion of the full Josephson cycle. Instead, the charging protocol can be optimized by interrupting the phase ramp at an interaction-dependent extraction phase
\(
\phi_{\mathrm{opt}}
\),
thereby maximizing the stored extractable work while avoiding subsequent coherent backflow processes.
Overall, these results establish that interaction-induced coherent dynamics fundamentally modify the operational charging strategy of ABS quantum batteries. Strong interactions not only enhance the charging rate and achievable ergotropy, but also generate phase-selective charging windows that enable optimized partial-cycle extraction protocols beyond conventional full-cycle operation.
%%%%%%%%%%%%%%%%%%%%%%%%%%%%%%%%%%%%%%%%%%%%%%%%%%%%%%%%%%%%%%%%%%%%%%%%%%%%%%%%%%%%%%%

\subsection{Ergotropy landscape in phase--interaction space}

\begin{figure}[t]
\centering
\includegraphics[width=0.72\columnwidth]{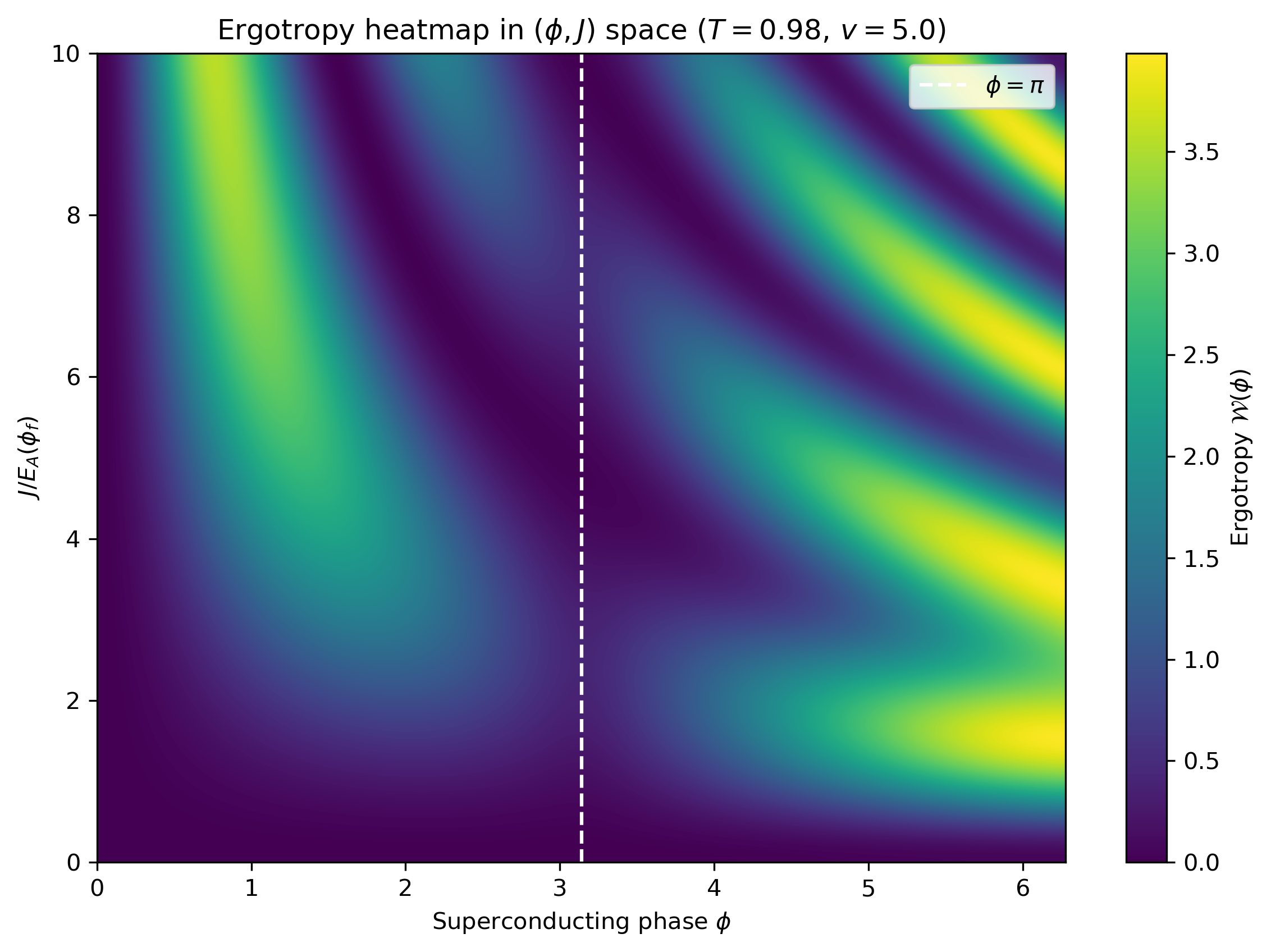}
\caption{
Phase-resolved ergotropy landscape $\mathcal{W}(\phi)$ in the combined superconducting phase--interaction parameter space for the high-transparency regime $T=0.98$ at fixed ramp speed $v=5.0$. The dashed vertical line marks the avoided crossing near $\phi=\pi$. 
}
\label{fig:ergotropy_landscape_phi_J}
\end{figure}

To obtain a global picture of the interaction-assisted charging dynamics, we examine the ergotropy landscape in the combined phase--interaction parameter space. Fig.~\ref{fig:ergotropy_landscape_phi_J} shows the phase-resolved ergotropy $\mathcal{W}(\phi)$ as a function of the superconducting phase $\phi$ and the scaled interaction strength $J/E_A(\phi_f)$ in the high-transparency regime $T=0.98$, where the avoided-crossing gap is strongly reduced and the ABS spectrum becomes highly phase sensitive.

Several important features emerge from the heatmap. In the weak-interaction regime, the ergotropy remains relatively small throughout the initial part of the Josephson cycle and only begins to increase significantly after the avoided crossing near $\phi = \pi$. This demonstrates that the dominant charging mechanism originates from nonadiabatic excitation across the minimum ABS gap. Before the avoided crossing, the phase evolution is comparatively adiabatic and only weakly populates non-passive states, leading to negligible extractable work.

As the interaction strength increases, the charging dynamics become progressively more structured and oscillatory. In particular, the heatmap develops pronounced diagonal high-ergotropy bands extending across the post-crossing region $\phi>\pi$. These oscillatory structures are direct signatures of coherent interaction-assisted redistribution dynamics between the coupled ABS sectors. Physically, the avoided crossing injects excitation into the system, while the inter-junction interaction hybridizes the ABS sectors and redistributes the excitation energy among different many-body configurations. The competition between these two mechanisms generates additional coherent timescales during the Josephson evolution.

The diagonal orientation of the bright oscillatory bands indicates that the optimal charging phase depends strongly on the interaction strength. As $J/E_A(\phi_f)$ increases, the phase at which maximal ergotropy is achieved shifts continuously during the cycle, revealing the emergence of interaction-controlled charging windows. These structures correspond to alternating constructive and destructive interference processes in the coherent redistribution dynamics. Regions of enhanced ergotropy arise when the phase-driven excitation and interaction-induced hybridization cooperate constructively, whereas the darker regions correspond to coherent backflow processes that partially suppress the extractable work.

Another important feature visible in the heatmap is the strong enhancement of the ergotropy amplitude at intermediate and large interaction strengths. While weak coupling produces only modest energy storage near to the final phase, larger interactions generate extended bright regions associated with highly non-passive states and efficient coherent redistribution before the final phase. 

The heatmap further demonstrates that the charging process is intrinsically phase selective. The largest extractable work is often achieved at phases substantially smaller than $2\pi$, particularly in the strongly interacting regime. This observation provides direct evidence that completion of the full Josephson cycle is not necessarily the optimal operational strategy for interacting ABS quantum batteries. Instead, the charging protocol can be optimized through interaction-dependent partial-cycle extraction, where the superconducting phase evolution is interrupted near the bright high-ergotropy regions before coherent backflow reduces the stored work.

Overall, the ergotropy landscape reveals that the charging dynamics of interacting ABS quantum batteries are governed by the interplay between avoided-crossing excitation, coherent hybridization, and phase-controlled interference effects. The figure therefore provides a direct visualization of phase-programmable quantum energy storage in superconducting hybrid systems.

%%%%%%%%%%%%%%%%%%%%%%%%%%%%%%%%%%%%%%%%%%%%%%%%%%%%%%%%%%%%

\section{Conclusion}

We have investigated the charging dynamics of a phase-driven quantum battery composed of two coupled Andreev bound state (ABS) units. Our results demonstrate that the charging process is governed by the interplay between nonadiabatic excitation near the superconducting avoided crossing and interaction-induced hybridization between the ABS sectors.

We identified distinct interaction-assisted charging regimes ranging from weak-coupling enhancement to strongly oscillatory coherent dynamics at larger interaction strength. In the high-transparency regime relevant for graphene SNS junctions, the reduced avoided-crossing gap strongly enhances the phase sensitivity of the charging dynamics and produces large interaction-driven modulation of the stored extractable work.

The phase-resolved analysis further revealed that the charging process is intrinsically nonmonotonic. Strong interactions generate coherent redistribution dynamics that produce multiple transient charging maxima during a single Josephson cycle. As a result, the maximum extractable work can occur substantially before completion of the full phase ramp.

This demonstrates that interacting ABS quantum batteries behave as phase-programmable energy-storage devices whose optimal operating point depends on the interplay between superconducting phase evolution and coherent interaction dynamics. In particular, the optimal partial-cycle protocol can significantly outperform conventional full-cycle extraction strategies.

The ergotropy landscape in the combined phase--interaction parameter space further revealed the emergence of interaction-dependent charging windows and coherent interference structures during the Josephson evolution. The resulting phase-selective charging dynamics provide a global picture of how avoided-crossing excitation and interaction-induced redistribution cooperate to generate extractable work in the interacting ABS battery.

Overall, our results establish interaction-assisted avoided-crossing dynamics as a physically transparent microscopic mechanism for generating and controlling extractable quantum work in superconducting hybrid systems. These findings provide a theoretical framework for phase-controlled quantum energy storage in experimentally accessible SNS junction architectures.

\bibliographystyle{apsrev4-2}
\bibliography{ref}

\end{document}